\documentclass[arxiv]{melba}

\usepackage{mwe} 

\usepackage{amsmath,amsfonts}

\usepackage{float}
\usepackage[T1]{fontenc}
\usepackage{makecell}
\usepackage{tablefootnote}
\usepackage{url}

\melbaid{YYYY:NNN}  
\doi{10.59275/j.melba.2024-AAAA}
\melbaauthors{Haghiri et al.}  
\email{hamideh.haghiri@dkfz-heidelberg.de} 
\volume{2}
\firstpageno{1}  
\melbayear{2025}  
\datesubmitted{2025-04-30}  
\datepublished{yyyy-m2-d2}  

\melbaspecialissue{Medical Imaging with Deep Learning (MIDL) 2020}
\melbaspecialissueeditors{Marleen de Bruijne, Tal Arbel, Ismail Ben Ayed, Hervé Lombaert}

\ShortHeadings{A Hybrid AI-based and Rule-based Approach to DICOM De-identification}{Haghiri et al.}

\title{A Hybrid AI-based and Rule-based Approach to DICOM De-identification: A Solution for the MIDI-B Challenge}

\author{
 \firstname Hamideh \surname Haghiri\aff{1} (ORCID: \href{https://orcid.org/0009-0006-7308-4235}{0009-0006-7308-4235}),
  \firstname Rajesh \surname Baidya\aff{1} (ORCID: \href{https://orcid.org/0009-0003-5235-1769}{0009-0003-5235-1769}),
  \firstname Stefan \surname Dvoretskii\aff{1} (ORCID: \href{https://orcid.org/0000-0001-7769-0167}{0000-0001-7769-0167}),
  \firstname Klaus H.\ \surname Maier-Hein\aff{1,2} (ORCID: \href{https://orcid.org/0000-0002-6626-2463}{0000-0002-6626-2463}),
  \firstname Marco \surname Nolden\aff{1,2} (ORCID: \href{https://orcid.org/0000-0001-9629-0564}{0000-0001-9629-0564})
}

\affiliations{
    \num 1 \addr Division of Medical and Biological Informatics, German Cancer Research Center, Heidelberg, Germany \\
    \num 2 \addr Pattern Analysis and Learning Group, Department of Radiation Oncology, Heidelberg University Hospital, Heidelberg, Germany \\
}

\abstract{
Ensuring the de-identification of medical imaging data is a critical step in enabling safe data sharing. This paper presents a hybrid de-identification framework designed to process Digital Imaging and Communications in Medicine (DICOM) files . Our framework adopts a modified, pre-built rule-based component, updated with The Cancer Imaging Archive (TCIA)'s best practices guidelines, as outlined in DICOM PS 3.15, for improved performance. It incorporates PaddleOCR, a robust Optical Character Recognition (OCR) system for extracting text from images, and RoBERTa, a fine-tuned transformer-based model for identifying and removing Personally Identifiable Information (PII) and Protected Health Information (PHI). Initially, the transformer-based model and the  rule-based component  were integrated to process for both structured data and free text. However, this coarse-grained approach did not yield optimal results. To improve performance, we refined our approach by applying the transformer model exclusively to free text, while structured data was handled only by rule-based methods.  In this framework the DICOM validator {\em dciodvfy} was leveraged  to ensure the integrity of DICOM files after the deID process. Through iterative refinement, including the incorporation of custom rules and private tag handling, the framework achieved a de-identification accuracy of 99.91\% on the MIDI-B test dataset. The results demonstrate the effectiveness of combining rule-based compliance with AI-enabled adaptability in addressing the complex challenges of DICOM de-identification.
Our code is available at~\url{https://github.com/MIC-DKFZ/miccai2024_midi-b-submission}.}

\keywords{{DICOM De-identification, AI-based de-identification, Rule-based de-identification, Protected Health Information (PHI), Personally Identifiable Information, RoBERTa model, PaddleOCR}}

\begin{document}

\twocolumn[\maketitle]

\section{Introduction}	
     Imaging plays a crucial role in numerous areas of medical and laboratory research, as well as clinical practice~\citep{Daoud:2020}. The growing access to medical images presents significant opportunities for research and development in healthcare. However, protecting patient privacy remains critical. This is where medical image de-identification becomes important. It involves removing or obscuring Personally Identifiable Information (PII) and Protected Health Information (PHI) from medical images and their related metadata~\citep{Dicom_DeIdentification}. PII includes direct identifiers such as names and birth dates, while PHI encompasses additional health-related identifiers such as medical record numbers and diagnostic codes~\citep{NCI_Workshop}. By effectively de-identifying medical images, we can safely share data for research, algorithm development, and other valuable purposes without compromising patient confidentiality~\citep{Dicom_DeIdentification}.
This paper presents our approach to the Medical Image De-Identification Benchmark (MIDI-B) challenge. MIDI-B offers a platform for developing and evaluating state-of-the-art de-identification of Digital Imaging and Communications in Medicine (DICOM) files using a large-scale, diverse data set aimed at accelerating progress in this critical domain~\citep{Rutherford:2021}.

\section{Related Works}

De-identification of medical images is a crucial step in ensuring privacy and compliance with data protection regulations such as the Health Insurance Portability and Accountability Act (HIPAA) and the General Data Protection Regulation (GDPR). Existing de-identification processes can be broadly classified into two main categories: DICOM header de-identification and pixel-level PII and PHI removal~\citep{shahid2022two}. To prevent re-identification, these attributes must be de-identified while preserving data utility and compliance with DICOM standards \citep{NEMA}

From the literature, we identify three main approaches to develop de-identification methods:

\subsection{Rule-Based DICOM De-identification} Rule-based approaches have traditionally been the most widely used methods for de-identifying DICOM metadata. Free DICOM de-identification tools have been extensively analyzed in previous studies, such as Aryanto et al. \citep{Aryanto2022}, where ten noncommercial DICOM toolkits were evaluated for their effectiveness in removing PHI from DICOM headers. The study found that only five tools could fully de-identify required elements, with four requiring careful customization. This highlights the limitations of rule-based approaches in ensuring complete de-identification, as different tools apply varying de-identification profiles with inconsistent results. While rule-based methods offer high interpretability and compliance with DICOM standard, they often lack adaptability when dealing with unstructured or non-standard metadata. Additionally, they require ongoing maintenance to ensure compliance with evolving regulatory standards.
\subsection{AI-based De-identification}
Recent advancements in machine learning (ML) and natural language processing (NLP) have facilitated AI-driven de-identification techniques. A systematic review by \citep{kovavcevic2024} analyzed AI-based de-identification methods for electronic health records (EHRs) from 2010 to 2023, covering 69 studies. The findings indicate that ML and hybrid models dominate the field, while  rule-based approaches are less frequently used. The i2b2/UTHealth corpus, a widely recognized dataset, remains the most commonly used benchmark for de-identification research.
Earlier AI-based methods relied on feature engineering and hybrid rule-ML approaches, whereas recent advancements leverage attention-based neural networks, achieving F1-scores exceeding 98\% on certain datasets. However, these models face challenges in generalizing across diverse clinical domains without additional manually labeled training data, raising concerns about their reliability in real-world applications. The study emphasizes the importance of improving evaluation metrics and expanding training data diversity to enhance the robustness of AI-driven de-identification.
Beyond text-based de-identification, deep learning methods have been applied to pixel-level PHI removal, particularly for burned-in PHI (e.g., embedded text in medical images). \citep{Khosravi2023} demonstrated the effectiveness of object detection models for radiographic marker removal while preserving essential medical annotations. However, AI-based de-identification also introduces challenges such as false positives, lack of interpretability, and regulatory concerns regarding automated data manipulation.
\subsection{Hybrid AI-based and Rule-Based De-identification}
Hybrid approaches that combine rule-based and AI-driven techniques have been explored to enhance automation, accuracy, and regulatory compliance in de-identification. These approaches typically apply rule-based logic to structured metadata fields, while leveraging machine learning for unstructured or ambiguous cases. A recent example is MedDiSC, an open-source tool that integrates rule-based metadata de-identification, AI-based pixel data processing using Optical Character Recognition (OCR), and automated DICOM curation and segmentation \citep{MedDiSC2024}. In MedDiSC, AI is primarily applied to the image level to detect and remove burned-in text, while metadata de-identification is handled through predefined rule sets.
Our work adopts a hybrid approach that incorporates AI components for both metadata and pixel data de-identification, alongside rule-based mechanisms. This design reflects an effort to address unstructured information across multiple parts of the DICOM file, while maintaining alignment with existing standards. Further details are described in the following sections.

\section{Methods}
	\subsection{Overview}

Our framework adopts a hybrid design, integrating a modified prebuilt rule-based component, a robust OCR system, and a fine-tuned transformer-based model to handle both metadata and image content. It also incorporates a DICOM validator to ensure post-deidentification DICOM integrity.
During the validation phase, we explored several modifications to improve performance. Initially, we employed the DCMPS315Deidentifier, a modified prebuilt algorithm that adhered to the DICOM PS 3.15 confidentiality guidelines~\citep{NEMA}. However, due to suboptimal performance scores, we developed a new algorithm, DCMTCIADeidentifier, based on the HIPAA Privacy Rule Safe Harbor method (DICOM PS 3.15 compliant) as defined by The Cancer Imaging Archive (TCIA) best practices guidelines~\citep{TCIA}.
Another modification involved adjusting the application of components. At first, the transformer-based model and the rule-based component were jointly applied to both structured data and free text. However, this broad application did not yield optimal results. To address this, we refined the approach by restricting the transformer model to free text, while structured data were handled exclusively by rule-based methods. These adjustments resulted in notable performance gains.
The framework comprises several components, which will be detailed in the following section.

	\subsection{Components}
	The de-identification framework is implemented through a set of components, each responsible for specific tasks in the de-identification process (see Figure 1).

   	\begin{figure*}[t]
		\centering
		\includegraphics[width=0.8\linewidth]{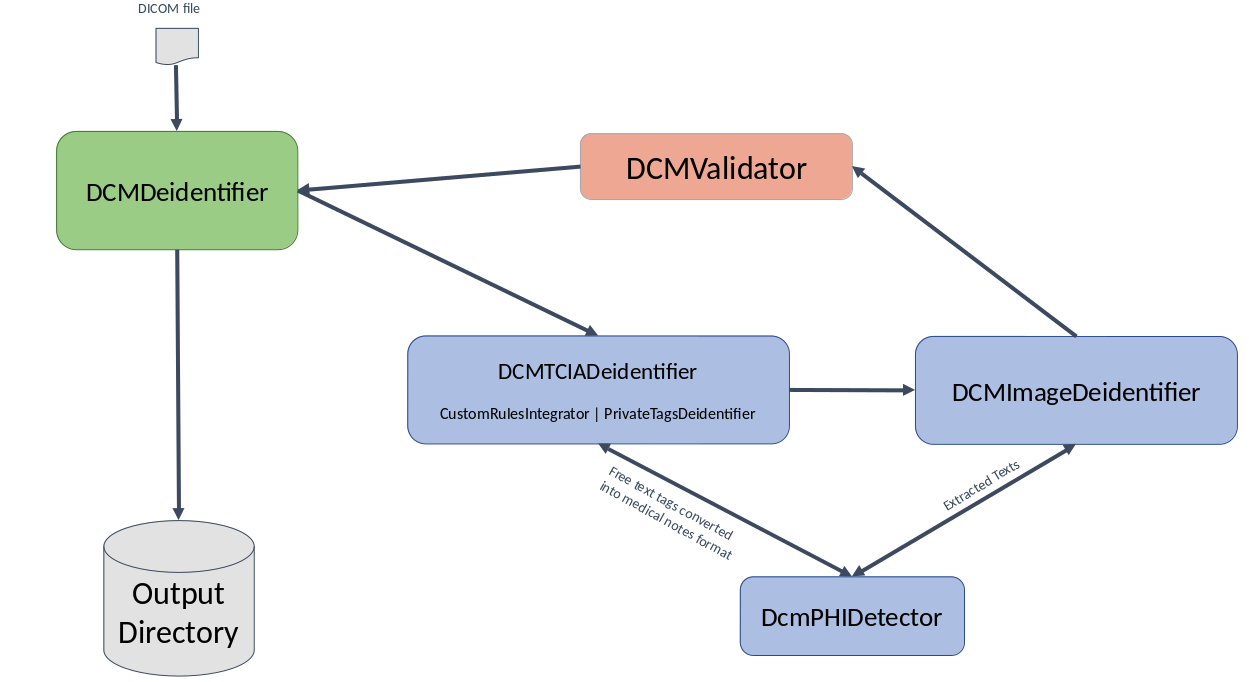}
		\caption{Core components of De-identification System.}
	\end{figure*} 
\subsubsection{\em DCMDeidentifier} 
This is the core component responsible for managing the entire de-identification workflow. It initializes all necessary de-identifiers and processes DICOM files across directories, and computes the de-identified output paths. The de-identification process is triggered through a ‘run’ function, which executes the following steps:

\begin{enumerate}[leftmargin=*, label=\textbf{Step \arabic*}:]
    \item De-identify the metadata of DICOM files.
    \item De-identify the image data by identifying and obscuring PHI detected within the images.
    \item Validate DICOM files.
    \item Export de-identified DICOM files to the appropriate directory.
\end{enumerate}

Additionally, the component exports a mapping of patient IDs and UIDs to a CSV file for reference and audit purposes.
\subsubsection{\em DcmPHIDetector}
In this component, due to limited resources for developing a custom AI model from scratch, obi/deid\_roberta\_i2b2, a fine-tuned RoBERTa model trained on the I2B2 2014 dataset was employed~\citep{OBI}. The RoBERTa model has been specifically fine-tuned for the de-identification of clinical data, leveraging the I2B2 2014 dataset, a benchmark dataset for medical text de-identification~\citep{stubbs2015annotating}. RoBERTa, an advanced Natural Language Processing (NLP) model, excels in contextual understanding, enabling it to accurately identify and classify PHI entities such as patient names, dates, and addresses~\citep{liu_roberta_2019-1}. By detecting and removing these sensitive data points, the model ensures compliance with privacy regulations such as HIPAA, while preserving the utility of the medical data for research and analysis~\citep{Emily2019}. The model’s primary functions include:
\begin{itemize}
    \item Identifies PHI entities in the text. \texttt{detect\_entities}
    \item Removes or obscures the detected PHI entities from the DICOM tag values. \texttt{deidentified\_element\_val}
\end{itemize}

\subsubsection{\em DCMPS315Deidentifier}
Baseline rule-based de-identification algorithm, built on top of  \texttt {dicomanonymizer.simpledicomanonymizer} ~\citep{GitHub} and follows the Application Level Confidentiality Profile Attributes described in DICOM PS31.5~\citep{NEMA}. It customizes core operation functions to incorporate the AI-based PHIDetector, enabling targeted de-identification of specific DICOM tags. Key features include:
Handling specific DICOM tags, such as those with Value Representations (VR) like "LO", "LT", "SH", "PN", "CS", "ST", "UT", where only PHI is replaced, leaving non-PHI text intact.
Shifting date-related tags (e.g., those with "DA", "DT" VR) by a configurable number of days (default is 120 days).
Ensuring consistency in patient IDs and UIDs by using mapping dictionaries.

\subsubsection{\em DCMTCIADeidentifier}
The {\em DCMTCIADeidentifier}, is like {\em DCMPS315Deidentifier}, but with the PS31.5 rules replaced by the rules from de-identification described by TCIA using HIPAA Privacy Rule Safe Harbor Method (DICOM PS 3.15 compliant)~\citep{TCIA}. All available private tags are excluded and left unmodified for processing by the private tags de-identifier component. The use of the AI PHI detector is minimized for efficiency. Tags with the action rules "replace" and "empty", as well as those with Value Representations (VR) such as "LO", "ST", and "LT" are treated as free text. These tags are consolidated into a single medical note format. The PHI detector is then applied to this combined document, and any identified sensitive information is replaced with an empty string to ensure privacy.

\subsubsection{\em DCMTCIADeidentifier with Custom Rules}
In addition to the Safe Harbor Method table, TCIA outlines several custom rules for de-identification. These include the removal of sensitive information such as allergies, patient state, occupation, and all comment fields, while certain technical identifiers such as station name, device serial number, device UID, and plate ID must be retained. Overall, 40 custom rules were implemented based on these guidelines to enhance the precision and compliance of the de-identification process.

\subsubsection{\em PrivateTagsDeidentifier}
This component utilizes a private tags dictionary provided by TCIA, containing 8,788 entries, to establish processing rules for private tags.  Each private tag is evaluated based on its associated value representation and the private block in which it resides. These elements are combined to formulate a key and specific processing rule for each tag in the private tags dictionary. For example, a sample private tag dictionary key might be formatted as "(0009,gems\_petd\_01,2b)\_SL," where \{tag group\}, \{private block\}, and \{last 2 digits of tag element\} are linked to a specific \{VR\}.
No AI-based PHI detection is applied during the de-identification of private tags.

\subsubsection{\em DCMImageDeidentifier}
This component is responsible for the de-identification of image data within DICOM files. It employs the PaddleOCR tool \citep{GitHubOCR} to extract text from DICOM slices. PaddleOCR is a comprehensive OCR system that supports various text detection and recognition tasks. It is known for its accuracy, speed, and ease of use, making it suitable for processing large volumes of images, such as DICOM slices in medical imaging. The tool is built on deep learning techniques and can handle a wide range of languages and text orientations, which is particularly important in medical imaging, where text may not always be aligned in a standard way. The extracted text is then processed by the {\em PHIDetector}, and if any PHI is detected, the corresponding bounding boxes within the images are filled with color from the neighboring pixel to obscure the sensitive information.

\subsubsection{\em DCMValidator}
This component employs  \texttt{dciodvfy} \citep{DICOMValidator} to ensure the integrity of DICOM files after the de-identification process. If validation errors arise due to missing attributes, the corresponding tags are added to the DICOM file with an empty value to maintain compliance and completeness.

\subsection{Challenges and Solutions}

During the validation phase, our PHI detector occasionally generated false positives. For instance, terms such as “MR BREAST” were frequently misclassified as person's name.  To mitigate such errors, we implemented a whitelist-based filtering strategy. Specifically, we compiled a list of 150 imaging-related terms—including “CT,” “MR,” “T1,” “abdomen,” and others—based on labeled anatomical case names extracted from an article by \cite{LabeledImagingAnatomy:2018}. Detected entities overlapping with this curated vocabulary were excluded from the final PHI detection, thereby reducing false positive rates without compromising the sensitivity of the de-identification process.

Additionally, after evaluating the performance on the validation data we noticed some of the "Missing Attributes" errors by the DCMValidator component were not relevant to the validation output. E.g. 'CodeValue', 'Manufacturer', 'ClinicalTrialSubjectID'. We decided to maintain an ignore list of such Missing Attribute tags and choose to ignore them while populating such tags after DICOM validation. 

\section{Results}
We applied our de-identification algorithms to both the MIDI-B validation dataset (Table \ref{table:1}) and MIDI-B test dataset (Table \ref{table:2}).The initial validation of the {\em DCMPS315Deidentifier} combined with the image de-identification module and use of AI component for all data achieved an accuracy of 84.36\%. The {\em DCMTCIADeidentifier} combined with the image de-identification module and selective use of AI component for free text, achieved 94.71\% accuracy. By incorporating custom rules, the accuracy improved to 97.65\%. Adding private tag de-identification further increased accuracy to 99.25\%. Implementing a second version of custom rules resulted in an accuracy of 99.88\%. Finally, integrating the {\em DcmValidator} tool with the latest version of custom rules and private tag de-identification achieved near-perfect accuracy of 99.98\%. The method was tested on an unseen test dataset, achieving an accuracy of 99.91\%, demonstrating acceptable performance.

\begin{table}[ht!]
    \caption{De-Identification Performance of Methods During Validation Phase}
    \label{table:1}
    \centering
    \renewcommand{\arraystretch}{1.3}
    \small 
    \resizebox{\linewidth}{!}{%
    \begin{tabular}{p{0.36
    \linewidth} p{0.40\linewidth} p{0.15\linewidth}}
    \toprule
    \textbf{Base Method\footnotemark[1]} & \textbf{Additional Processing} & \textbf{Score} \\
    \midrule
    DCMPS315DeIdentifier & & 84.36\% \\
    \midrule
    DCMTCIADeIdentifier & & 94.71\% \\
    \midrule
    DCMTCIADeIdentifier & CustomRules & 97.65\% \\
    \midrule
    DCMTICADeIdentifier & \makecell[l]{CustomRules\\PrivateTagsDeIdentifier} & 99.25\% \\
    \midrule
    DCMTCIADeIdentifier & \makecell[l]{CustomRulesV2\\PrivateTagsDeIdentifier} & 99.88\% \\
    \midrule
    DCMTCIADeIdentifier & \makecell[l]{CustomRulesV2\\PrivateTagsDeIdentifier\\ DcmValidator} & 99.98\% \footnotemark[2] \\
    \bottomrule
    \end{tabular}
    }
\end{table}
\noindent
\begin{table}[ht!]
    \caption{De-Identification Performance of Final Method on Test Phase}
    \label{table:2}
    \centering
    \renewcommand{\arraystretch}{1.3} 
    \small
    \resizebox{\linewidth}{!}{%
        \begin{tabular}{p{0.36
    \linewidth} p{0.40\linewidth} p{0.15\linewidth}}
            \toprule
            \textbf{Base Method\footnotemark[1]} & \textbf{Addictional Processing} & \textbf{Score} \\
            \midrule
            DCMTCIADeIdentifier & \makecell[l]{CustomRulesV2\\PrivateTagsDeIdentifier\\DcmValidator} & 99.91\% \\
            \bottomrule
        \end{tabular}
    }
\end{table}

\footnotetext[1]{All the methods also use DCMImageDeIdentifier with the base methods}
\footnotetext[2]{This score was computed using a self-derived method based on the difference from the previous validation score after the validation phase.}

\begin{table}[t]
    \caption{Test Output Discrepancy Report Analysis}
    \label{table:3}
    \vspace{1em}
    \centering
    \small
    \renewcommand{\arraystretch}{1.3} 
    \resizebox{\linewidth}{!}{%
    \begin{tabular}{p{0.45\linewidth} p{0.1\linewidth} p{0.45\linewidth} p{0.1\linewidth}}
    \toprule
    \textbf{Tag Name} & \textbf{Failed} & \textbf{Tag Name} & \textbf{Failed}\\
    \midrule
    \texttt{<Protocol Name>} & 962 & \texttt{<Study Comments>} & 132 \\
    \texttt{<Series Description>} & 908 & \texttt{<Performed Procedure Step Description>} & 117 \\
    \texttt{<Study Description>} & 709 & \texttt{<Image Type>} & 101 \\
    \texttt{<Medical Alerts>} & 661 & \texttt{<Referenced SOP Class UID>} & 96 \\
    \texttt{<Additional Patient History>} & 510 & \texttt{<Referenced File ID>} & 64 \\
    \texttt{<Private Creator>} & 409 & \texttt{<Presentation LUT Shape>} & 39 \\
    \texttt{<Requested Procedure Description>} & 278 & \texttt{<Presentation Intent Type>} & 36 \\
    \texttt{<Derivation Description>} & 256 & \texttt{<Comments on the Performed Procedure Step>} & 32 \\
    \texttt{<Contrast/Bolus Agent>} & 171 & \multicolumn{2}{c}{} \\ 
    \bottomrule
    \end{tabular}
    }
\end{table}
\noindent
\begin{table}[t]
    \caption{Runtime}
    \label{table:4}
    \centering
    \small
    \vspace{1em}
    \renewcommand{\arraystretch}{1.3} 
    \resizebox{\linewidth}{!}{%
    \begin{tabular}{p{0.28\linewidth} p{0.18\linewidth} p{0.38\linewidth} p{0.16\linewidth}}
    \toprule
    \textbf{Dataset} & \textbf{DICOM Files} & \textbf{Algorithm} & \textbf{Runtime} \\
    \midrule
    Validation Dataset & 23,921 & DCMPS315DeIdentifier & 1:10:41 \\
    Validation Dataset & 23,921 & DCMTCIADeIdentifier & 1:08:44 \\
    Validation Dataset & 23,921 & DCMTCIADeIdentifier \& DICOM Validator & 2:04:36 \\
    Test Dataset & 29,660 & DCMTCIADeIdentifier & 1:32:29 \\
    Test Dataset & 29,660 & DCMTCIADeIdentifier \& DICOM Validator & 2:33:50 \\
    \bottomrule
    \end{tabular}
    }
\end{table}

Table \ref{table:3} presents the analysis of discrepancies found in the test output for various DICOM tags after the de-identification process. It lists specific tag names and the number of instances (failed) where discrepancies occurred. \\

All experiments were conducted on a machine equipped with an AMD Ryzen 9 3900X 12-Core Processor, 64 GB of RAM, and a GPU with 12 GB of VRAM. The runtimes of our de-identification algorithms are presented on Table \ref{table:4}. \\

\begin{figure}[h]
    \centering
    \includegraphics[width=\linewidth, trim=0cm 0cm 0cm 2cm, clip]{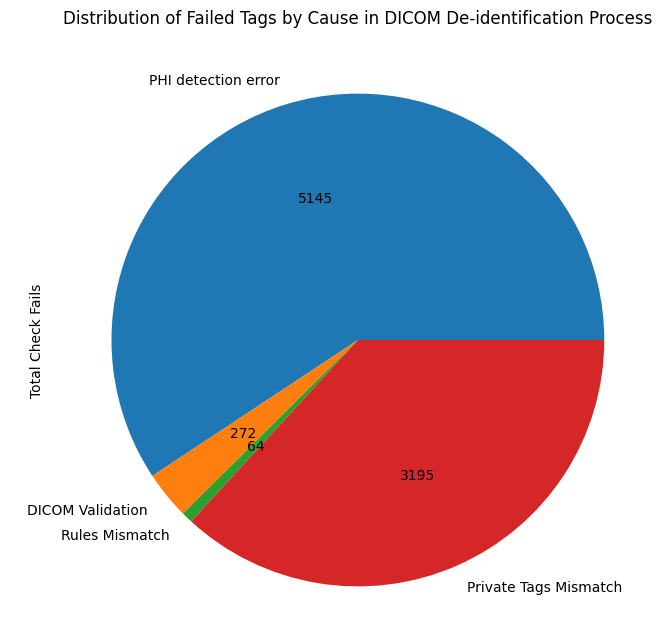}
    \caption{De-identification Failures by Category}
    \label{fig:fig1}
\end{figure}

To gain insights into the failures of our de-identification algorithm, we analyzed all of the 8,676 mismatches from the test data that are reported in the discrepancy report generated by the MIDI-B validation script. The first pie chart (Figure: \ref{fig:fig1}) illustrates the distribution of these failures across four distinct categories: the majority (5,145 cases, 59.3\%) stemmed from limitations in PHI detection, indicating missed identification of sensitive information. A substantial number of failures (3,195 cases, 36.8\%) were attributed to incorrect handling of private DICOM tags, highlighting the need for a more robust and comprehensive rule set tailored to these vendor-specific elements. An additional 272 mismatches (3.1\%) resulted from inconsistencies flagged during DICOM validation, and 64 failures (0.7\%) were caused by incomplete or inappropriately applied custom rules.

\begin{figure}[h]
    \includegraphics[width=\linewidth, trim=0cm 0cm 0cm 2cm, clip]{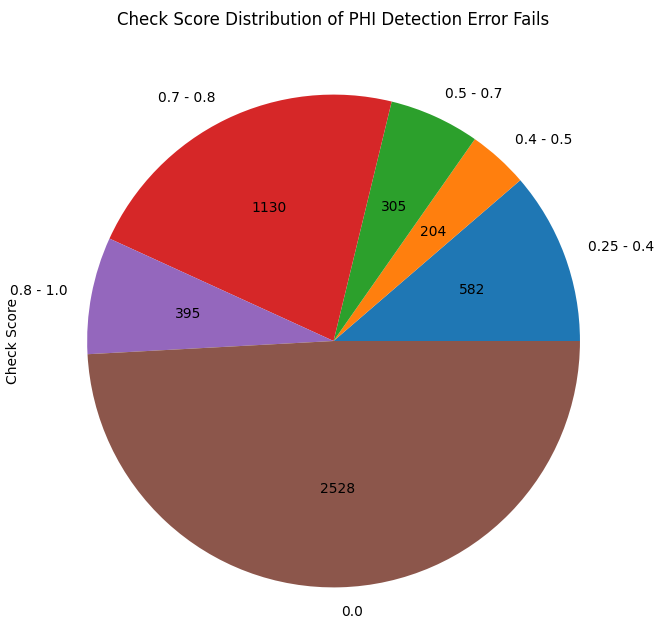}
    \caption{Check Score Distribution}
    \label{fig:fig2}
\end{figure}

The second pie chart (Figure: \ref{fig:fig2}) presents the distribution of the \texttt{check\_score} values associated with these mismatched tags, a metric that reflects the degree of similarity between the expected and the algorithm's de-identified output. A total of 2,528 tags (49.1\%) received a check score of 0.0, indicating a complete mismatch. Scores in the range of 0.25–0.4 accounted for 582 tags (11.3\%), while 204 tags (3.9\%) fell within the 0.4–0.5 range. Moderate similarity (check scores between 0.5–0.7) was observed in 305 tags (5.9\%), and 1,130 tags (21.9\%) achieved partial alignment with scores between 0.7–0.8. Notably, 395 tags (7.6\%) achieved high similarity (0.8–0.99), suggesting minor deviations from the expected output. These findings underscore the varying degrees of error severity and offer targeted directions for improving both PHI detection and content-preserving transformation strategies.


\section{Discussion}

The development of a robust de-identification framework for DICOM files presents unique challenges, particularly in balancing patient privacy with data utility. Our approach to the MIDI-B challenge highlights both the strengths and limitations of relying on transformer-based models for PHI detection. While AI-based methods such as RoBERTa demonstrate strong performance in identifying PHI in free-text contexts, their effectiveness diminishes when applied to structured metadata fields common in DICOM files. These cases often require complementary rule-based logic to ensure accurate de-identification. The iterative refinements applied throughout the challenge led to progressively improved results, underscoring the importance of a hybrid, adaptive approach. Our final model achieved an exceptionally high score of 99.91\%, indicating near-perfect de-identification on the provided test set.

However, despite this strong performance, it is important to acknowledge that this score may not generalize across all data sources or clinical settings. DICOM metadata varies widely depending on modality, institution, and vendor-specific implementations. Therefore, evaluating de-identification approaches across diverse datasets remains a critical challenge. 

Future work could explore fine-tuning PHI detection models specifically on DICOM metadata rather than relying on models that are trained on general clinical data. Such models, trained on DICOM-specific formats and semantics, may better distinguish between sensitive information and domain-specific terminology, thereby improving generalizability and robustness across varied use cases.

\acks{This work was supported by the Helmholtz Metadata Collaboration (HMC) Hub Health, and by the RACOON project in „NUM 2.0" (FKZ: 01KX2121), BMBF.}

%
\ethics{The work follows appropriate ethical standards in conducting research and writing the manuscript, following all applicable laws and regulations regarding treatment of animals or human subjects.}

\coi{The authors have no competing interests to declare that are relevant to the content of this article. }

\data{Our code is available at~\url{https://github.com/MIC-DKFZ/miccai2024_midi-b-submission}
}

\bibliography{sample}


\end{document}